\input harvmac

%
\let\includefigures=\iftrue
%
%
%
\newfam\black
\input rotate
\input epsf
\noblackbox
%
%
\includefigures
\message{If you do not have epsf.tex (to include figures),}
\message{change the option at the top of the tex file.}
\def\figin{\epsfcheck\figin}\def\figins{\epsfcheck\figins}
\def\epsfcheck{\ifx\epsfbox\UnDeFiNeD
\message{(NO epsf.tex, FIGURES WILL BE IGNORED)}
\gdef\figin##1{\vskip2in}\gdef\figins##1{\hskip.5in}
\else\message{(FIGURES WILL BE INCLUDED)}%
\gdef\figin##1{##1}\gdef\figins##1{##1}\fi}
\def\DefWarn#1{}

\def\figinsert{\goodbreak\midinsert}
\def\ifig#1#2#3{\DefWarn#1\xdef#1{fig.~\the\figno}
\writedef{#1\leftbracket fig.\noexpand~\the\figno}%
\figinsert\figin{\centerline{#3}}\medskip\centerline{\vbox{\baselineskip12pt
\advance\hsize by -1truein\noindent\footnotefont{\bf
Fig.~\the\figno:} #2}}
\bigskip\endinsert\global\advance\figno by1}
\else
\def\ifig#1#2#3{\xdef#1{fig.~\the\figno}
\writedef{#1\leftbracket fig.\noexpand~\the\figno}%
\global\advance\figno by1} \fi
\def\yboxit#1#2{\vbox{\hrule height #1 \hbox{\vrule width #1
\vbox{#2}\vrule width #1 }\hrule height #1 }}
\def\fillbox#1{\hbox to #1{\vbox to #1{\vfil}\hfil}}
\def\ybox{{\lower 1.3pt \yboxit{0.4pt}{\fillbox{8pt}}\hskip-0.2pt}}

\def\rightarrowbox#1#2{
  \setbox1=\hbox{\kern#1{${ #2}$}\kern#1}
  \,\vbox{\offinterlineskip\hbox to\wd1{\hfil\copy1\hfil}
    \kern 3pt\hbox to\wd1{\rightarrowfill}}}

\def\p{\partial}

\def\Tr{{{\rm Tr~ }}}

\def\CL{{\cal L}}

\def\CN{{\cal N}}

\def\tilde{\widetilde}

\def\II{\relax{I\kern-.10em I}}

\def\bar{\overline}

\def\IZ{\relax\ifmmode\mathchoice
{\hbox{\cmss Z\kern-.4em Z}}{\hbox{\cmss Z\kern-.4em Z}}
{\lower.9pt\hbox{\cmsss Z\kern-.4em Z}} {\lower1.2pt\hbox{\cmsss
Z\kern-.4em Z}}\else{\cmss Z\kern-.4em Z}\fi}
\def\IB{\relax{\rm I\kern-.18em B}}
\def\IC{{\relax\hbox{$\inbar\kern-.3em{\rm C}$}}}
\def\ID{\relax{\rm I\kern-.18em D}}
\def\IE{\relax{\rm I\kern-.18em E}}
\def\IF{\relax{\rm I\kern-.18em F}}
\def\IG{\relax\hbox{$\inbar\kern-.3em{\rm G}$}}
\def\IGa{\relax\hbox{${\rm I}\kern-.18em\Gamma$}}
\def\IH{\relax{\rm I\kern-.18em H}}
\def\II{\relax{\rm I\kern-.18em I}}
\def\IK{\relax{\rm I\kern-.18em K}}
\def\IN{\relax{\rm I\kern-.18em N}}
\def\IP{\relax{\rm I\kern-.18em P}}

%
\def\inbar{\,\vrule height1.5ex width.4pt depth0pt}

\def\p{\partial}

\font\cmss=cmss10 \font\cmsss=cmss10 at 7pt
\def\IR{\relax{\rm I\kern-.18em R}}

\def\lp10{l_P^{10}}
\def\lp11{l_P^{11}}
\def\R11{R_{11}}

\def\alphadot{{\dot \alpha}}
\def\betadot{{\dot \beta}}
\def\gammadot{{\dot \gamma}}

\def\thetabar{{\bar\theta}}

\def\tilde{\widetilde}
\def\CN{{\cal N}}

%
%
\lref\WessCP{ J.~Wess and J.~Bagger, ``Supersymmetry And
Supergravity,'' Princeton, USA: Univ. Pr. (1992).
}

\lref\BrittoAJ{ R.~Britto, B.~Feng and S.~J.~Rey, ``Deformed
Superspace, N=1/2 Supersymmetry and (Non)Renormalization
Theorems,'' arXiv:hep-th/0306215.
}

\lref\SchwarzPF{ J.~H.~Schwarz and P.~Van Nieuwenhuizen,
``Speculations Concerning A Fermionic Substructure Of
Space-Time,'' Lett.\ Nuovo Cim.\  {\bf 34}, 21 (1982).
}

\lref\FradkinFT{ E.~S.~Fradkin and A.~A.~Tseytlin, ``Non-linear
electrodynamics from quantized strings'', Phys. Lett. {\bf B163},
123 (1985).
}

\lref\YostYY{ A.~Abouelsaood, C.~Callan, C.~Nappi and S.~Yost, ``Open
strings in background gauge fields,'' Nucl. Phys. {\bf B280},
599 (1987).
}

\lref\PershinBF{ N.~Berkovits and V.~Pershin, ``Supersymmetric
Born-Infeld from the pure spinor formalism of the open
superstring,'' JHEP {\bf 0301}, 023 (2003)
[arXiv:hep-th/0205154].
}

\lref\BaggerPI{ J.~Bagger and A.~Galperin, ``The tensor Goldstone
multiplet for partially broken supersymmetry,'' Phys.\ Lett.\ B
{\bf 412}, 296 (1997) [arXiv:hep-th/9707061].
}

\lref\BaggerWP{ J.~Bagger and A.~Galperin, ``A new Goldstone
multiplet for partially broken supersymmetry,'' Phys.\ Rev.\ D
{\bf 55}, 1091 (1997) [arXiv:hep-th/9608177].
}

\lref\BerkovitsBF{ N.~Berkovits, ``A new description of the
superstring,'' arXiv:hep-th/9604123.
}

\lref\BershadskyBF{ N.~Berkovits, M.~Bershadsky, T.~Hauer,
S.~Zhukov and B.~Zwiebach, ``Superstring theory on
$AdS_2\times S^2$ as a coset supermanifold,'' Nucl. Phys.
{\bf B567}, 61 (2000).[arXiv:hep-th/9907200].
}

\lref\BrinkNZ{ L.~Brink and J.~H.~Schwarz, ``Clifford Algebra
Superspace,'' CALT-68-813
{SPIRES entry}
}

\lref\SeibergVS{ N.~Seiberg and E.~Witten, ``String theory and
noncommutative geometry,'' JHEP {\bf 9909}, 032 (1999)
[arXiv:hep-th/9908142].
}

\lref\KlemmYU{ D.~Klemm, S.~Penati and L.~Tamassia,
``Non(anti)commutative superspace,'' arXiv:hep-th/0104190.
}

\lref\ChepelevGA{ I.~Chepelev and C.~Ciocarlie, ``A path integral
approach to noncommutative superspace,'' arXiv:hep-th/0304118.
}

\lref\deBoerDN{ J.~de Boer, P.~A.~Grassi and P.~van Nieuwenhuizen,
``Non-commutative superspace from string theory,''
arXiv:hep-th/0302078.
}

\lref\KawaiYF{ H.~Kawai, T.~Kuroki and T.~Morita, ``Dijkgraaf-Vafa
theory as large-N reduction,'' arXiv:hep-th/0303210.
}

\lref\CasalbuoniHX{ R.~Casalbuoni, ``Relativity And
Supersymmetries,'' Phys.\ Lett.\ B {\bf 62}, 49 (1976).
}

\lref\CasalbuoniBJ{ R.~Casalbuoni, ``On The Quantization Of
Systems With Anticommutating Variables,'' Nuovo Cim.\ A {\bf 33},
115 (1976).
}

\lref\CasalbuoniTZ{ R.~Casalbuoni, ``The Classical Mechanics For
Bose-Fermi Systems,'' Nuovo Cim.\ A {\bf 33}, 389 (1976).
}

\lref\OoguriQP{ H.~Ooguri and C.~Vafa, ``The C-deformation of
gluino and non-planar diagrams,'' arXiv:hep-th/0302109.
}

\lref\OoguriTT{ H.~Ooguri and C.~Vafa, ``Gravity induced
C-deformation,'' arXiv:hep-th/0303063.
}

\lref\AbbaspurXJ{ R.~Abbaspur, ``Generalized noncommutative
supersymmetry from a new gauge symmetry,'' arXiv:hep-th/0206170.
}

\lref\FerraraMM{ S.~Ferrara and M.~A.~Lledo, ``Some aspects of
deformations of supersymmetric field theories,'' JHEP {\bf 0005},
008 (2000) [arXiv:hep-th/0002084].
}

\lref\DouglasBA{ M.~R.~Douglas and N.~A.~Nekrasov,
``Noncommutative field theory,'' Rev.\ Mod.\ Phys.\  {\bf 73}, 977
(2001) [arXiv:hep-th/0106048].
}

\lref\SeibergYZ{ N.~Seiberg, ``Noncommutative superspace, N = 1/2
supersymmetry, field theory and  string theory,'' JHEP {\bf 0306},
010 (2003) [arXiv:hep-th/0305248].
}

\lref\DavidKE{ J.~R.~David, E.~Gava and K.~S.~Narain, ``Konishi
anomaly approach to gravitational F-terms,'' arXiv:hep-th/0304227.
}

\noblackbox

\newbox\tmpbox\setbox\tmpbox\hbox{\abstractfont
}
 \Title{\vbox{\baselineskip12pt\hbox to\wd\tmpbox{\hss
 hep-th/0306226} }}
 {\vbox{\centerline{Superstrings in Graviphoton Background}
 \centerline{}
 \centerline{and
 $\CN={1\over 2} + {3\over 2}$ Supersymmetry}}
 }
\smallskip
\centerline{Nathan Berkovits}
\bigskip
\centerline{Instituto de F\'{\i}sica Te\'orica, Universidade
Estadual Paulista} \centerline{R. Pamplona 145, S\~ao Paulo, SP
01405-900 BRASIL }
\bigskip
\centerline{Nathan Seiberg}
\bigskip
\centerline{School of Natural Sciences, Institute for Advanced
Study}
\centerline{Princeton NJ 08540 USA}
\bigskip
\vskip 1cm
 \noindent
Motivated by Ooguri and Vafa, we study superstrings in flat
$\IR^4$ in a constant self-dual graviphoton background. The
supergravity equations of motion are satisfied in this background
which deforms the $\CN=2$ $d=4$ flat space super-Poincar\'e
algebra to another algebra with eight supercharges.  A $D$-brane
in this space preserves a quarter of the supercharges; i.e.\
$\CN={1\over 2} $ supersymmetry is realized linearly, and the
remaining $\CN={3\over 2}$ supersymmetry is realized nonlinearly.
The theory on the brane can be described as a theory in
noncommutative superspace in which the chiral fermionic
coordinates $\theta^\alpha$ of $\CN=1$ $d=4$ superspace are not
Grassman variables but satisfy a Clifford algebra.

\Date{June 2003}
%

%
%

\newsec{Introduction}

Motivated by the work of Ooguri and Vafa \OoguriQP\ we continue
the analysis of \SeibergYZ\ of superstrings in flat $\IR^4$ with
$\CN=2$ supersymmetry with constant self-dual graviphoton field
strength.

Only in Euclidean space can we turn on the self-dual part of the
two form field strength $F^{\alpha\beta}$ while setting the
anti-self-dual part $F^{\alphadot\betadot}$ to zero.  Therefore we
limit the discussion to Euclidean space, but we will use
Lorentzian signature notation\foot{We will use the conventions of
Wess and Bagger \WessCP.}.

The special property of a purely self-dual field strength
$F^{\alpha\beta}$ is that it does not contribute to the energy
momentum tensor, and therefore does not lead to a source in the
gravity equations of motion.  Also, since the kinetic term of the
graviphoton does not depend on the dilaton, the background
$F^{\alpha\beta}$ does not lead to a source in the dilaton
equation.  Therefore, this background is a solution of the
equations of motion.

Before turning on the graviphoton our system has $\CN=2$
supersymmetry. What happens to it as a result of the nonzero
$F^{\alpha\beta}$? The local $\CN=2$ $d=4$ chiral gravitini
$\xi_\mu^{ \beta j}$ transformation laws are
 \eqn\gravit{\delta \xi_\mu^{\beta j} =
 \sigma_{\mu\alpha\alphadot}\bar\varepsilon^{\alphadot j}
 \alpha' F^{\alpha\beta} + \nabla_\mu \varepsilon^{\beta j}}
where $\varepsilon^{\beta j}$ and $\varepsilon^{\betadot j}$ are
the local supersymmetry parameters. The dilatini and antichiral
gravitini do not transform into the self-dual graviphoton field
strength.  The variation \gravit\ clearly vanishes for constant
$\varepsilon^{\beta j}$ and $\bar\varepsilon^{\alphadot j}=0$.
This shows that the background $F^{\alpha\beta}$ does not affect
the four supercharges with one chirality $Q_{\alpha i}$.  The
variation \gravit\ is not zero for constant
$\bar\varepsilon^{\alphadot j}$ showing that the four other
standard supercharges are broken by $F^{\alpha\beta}$. However,
one can easily make the variation of the chiral gravitini \gravit\
vanish if one chooses
 \eqn\news{\varepsilon^{\beta j} = {1\over 2}\alpha' F^{\alpha\beta}
 \sigma_{\mu\alpha\alphadot} \bar\varepsilon^{\alphadot j} x^\mu}
where $\bar\varepsilon^{\alphadot j}$ is constant.  So in the
presence of this background, one has eight global fermionic
symmetries which generate a deformed $\CN=2$ $d=4$ supersymmetry.

One can easily read off the algebra of the deformed $\CN=2$ $d=4$
supersymmetry from the above transformations
 \eqn\newalg{\eqalign{
  &\{Q^\alpha_j, \bar Q^\alphadot_k\} = 2 \epsilon_{jk}
 \sigma_\mu^{\alpha\alphadot} P^\mu\cr
  &[P_\mu,  Q_{\alpha j}] = [P_\mu, P_\nu]
 = \{Q_{\alpha j},Q_{\beta k}\}= 0\cr
 &[P_\mu, \bar Q_{\alphadot j}] =  2 \sigma_{\mu\beta\alphadot}
 \alpha' F^{\alpha\beta} Q_{\alpha j} \cr
 &\{\bar Q_{\alphadot j},\bar Q_{\betadot k}\}= 4\epsilon_{jk}
 \epsilon_{\alphadot\betadot} \alpha'
 F^{\alpha\beta}M_{\alpha\beta}\cr
 }}
where
$M_{\alpha\beta}$ is the self-dual Lorentz generator which is
normalized to transform
 \eqn\Mtra{[M_{\alpha\beta},\theta^\gamma] =\theta_\alpha
 \delta_\beta^\gamma + \theta_\beta\delta_\alpha^\gamma .}

Note that if one had turned on both a self-dual field strength
$F^{\alpha\beta}$ and an anti-self-dual field strength
$F^{\alphadot\betadot}$, there would have been a backreaction
which would warp the spacetime to (Euclidean) $AdS_2\times S^2$.
The isometries of this space form a $PSU(2|2)$ algebra, and the
algebra of \newalg\ can be understood as a certain contraction of
$PSU(2|2)$. In \BershadskyBF, the superstring was studied in this
$AdS_2\times S^2$ background and the structure of the worldsheet
action closely resembles the pure spinor version of the
$AdS_5\times S^5$ worldsheet action. Since the worldsheet action
becomes quadratic in the limit where the anti-self-dual field
strength $F^{\alphadot\betadot}$ goes to zero, this limit might be
useful for studying $AdS_d\times S^d$ backgrounds in a manner
analogous to the Penrose limit.

In the next section we will extend this  supergravity discussion
to superstring theory. Using the hybrid formalism for the
superstring, the worldsheet action remains quadratic in this
graviphoton background, so the theory is trivial to analyze. We
will explicitly show that it preserves eight supercharges with the
new algebra \newalg.  In section 3 we will study $D$-branes in
this background, and will examine the symmetries on the branes. In
section 4 we will consider the low energy field theory on the
$D$-brane and will examine its supersymmetries.

\newsec{Superstrings in graviphoton background}

In this section and the next one we review and extend the
discussion in \refs{\OoguriQP,\SeibergYZ} of
type II superstrings in $\IR^4$ with $\CN=2$ supersymmetry
deformed by a self-dual graviphoton (see also \deBoerDN).

Since we are interested in superstrings in Ramond-Ramond background,
the standard NSR formalism cannot be used.  Instead, we will use
the hybrid formalism (for a review see \BerkovitsBF).  The
relevant part of the worldsheet Lagrangian is
 \eqn\wslagII{ \CL = {1\over \alpha'}\left({1\over 2} \tilde
 \partial x^\mu \partial x_\mu + p_\alpha \tilde \partial
 \theta^\alpha + \bar p_\alphadot \tilde
 \partial\thetabar^\alphadot + \tilde p_\alpha \partial
 \tilde\theta^\alpha + \tilde {\bar p}_\alphadot
 \partial\tilde\thetabar^\alphadot\right),}
where $\mu=0\dots 3$, $(\alpha,\alphadot)=1,2$, and we ignore the
worldsheet fields $\rho$ and the Calabi-Yau sector.  Since we use
a bar to denote the space time chirality, we use a tilde to denote
the worldsheet chirality.  Therefore when the worldsheet has
Euclidean signature it is parametrized by $z$ and $\tilde z$ which
are complex conjugate of each other.   $p$, $\tilde p$, $\bar p$
and $\tilde {\bar p}$ are canonically conjugate to $\theta$,
$\tilde \theta$ $\thetabar$ and $\tilde \thetabar$; they are the
worldsheet versions of $-{\partial \over\partial \theta} \big|_x$
etc..  The reason $x$ is held fixed in these derivatives is that
$x$ appears as another independent field in \wslagII.

It is convenient to change variables to
 \eqn\chav{\eqalign{
 y^\mu &=x^\mu+i \theta^\alpha\sigma^\mu_{\alpha\alphadot}
 \thetabar^\alphadot  +i \tilde \theta^\alpha\sigma^
 \mu_{\alpha\alphadot}\tilde\thetabar^\alphadot\cr
 \bar d_\alphadot &= \bar p_\alphadot -i \theta^\alpha
 \sigma^\mu_{\alpha\alphadot}\partial x_\mu - \theta\theta
 \partial\thetabar_\alphadot +{1\over 2} \thetabar_\alphadot
 \partial (\theta\theta)\cr
 q_\alpha  &=-p_\alpha -i \sigma^\mu_{\alpha\alphadot}
 \thetabar^\alphadot \partial x_\mu +{1\over 2}
 \thetabar\thetabar \partial\theta_\alpha -{3\over 2}\partial
 (\theta_\alpha \thetabar\thetabar) \cr
 \tilde{\bar d}_\alphadot &= \tilde{ \bar p}_\alphadot -i \tilde
 \theta^\alpha \sigma^\mu_{\alpha\alphadot}\tilde \partial x_\mu
 - \tilde\theta\tilde\theta \tilde \partial\tilde
 \thetabar_\alphadot +{1\over 2} \tilde\thetabar_\alphadot
  \tilde \partial (\tilde\theta\tilde\theta)\cr
 \tilde q_\alpha  &=-\tilde p_\alpha-i\sigma^\mu_{\alpha
 \alphadot}\tilde \thetabar^\alphadot \tilde \partial x_\mu
 +{1\over 2} \tilde\thetabar\tilde \thetabar \tilde \partial
 \tilde \theta_\alpha -{3\over 2}\tilde \partial (\tilde
 \theta_\alpha \tilde \thetabar\tilde \thetabar)}}
and to derive
 \eqn\wslagIIa{{\cal L}={1\over \alpha'}\left({1\over 2} \tilde
 \partial y^\mu \partial y_\mu -q_\alpha \tilde \partial
 \theta^\alpha
 + \bar d_\alphadot \tilde \partial\thetabar^\alphadot - \tilde
 q_\alpha \partial \tilde \theta^\alpha + \tilde{\bar d}_\alphadot
 \partial\tilde \thetabar^\alphadot + {\rm total
 ~derivative}\right). }

The new variables in \chav\ are the worldsheet versions of $\bar
D_\alphadot = -{\partial \over \partial \thetabar^\alphadot}
\big|_y$, $Q_\alpha = {\partial \over \partial \theta^\alpha}
\big|_y$, etc..  The reason $y$ is held fixed in these derivatives
is that it appears as an independent field in \wslagIIa.  Our
definitions of $q$ and $\tilde q$ differ from the integrand of the
supercharges in \BerkovitsBF\ by total derivatives which do not
affect the charges, but are important for our purpose.  We will
also need the worldsheet versions of $D_\alpha$, $\tilde D_\alpha$
and the other two supercharges
 \eqn\dqdef{\eqalign{
 d_\alpha &= -p_\alpha + i \sigma^\mu_{\alpha\alphadot}
  \thetabar^\alphadot \partial x_\mu -\thetabar\thetabar
 \partial\theta_\alpha +{1\over 2} \theta_\alpha \partial
 (\thetabar\thetabar)= q_\alpha + 2i \sigma^\mu_{\alpha\alphadot}
 \thetabar^\alphadot \partial y_\mu - 4 \thetabar\thetabar
 \partial \theta_\alpha\cr
  \tilde d_\alpha &= -\tilde p_\alpha + i  \sigma^\mu_
  {\alpha\alphadot} \thetabar^\alphadot \tilde\partial x_\mu
  -\tilde\thetabar \tilde\thetabar \tilde
  \partial\tilde\theta_\alpha  +{1\over 2}  \tilde\theta_\alpha
  \tilde \partial (\tilde\thetabar\tilde\thetabar)=
  \tilde q_\alpha + 2i\sigma^\mu_{\alpha\alphadot}
 \tilde \thetabar^\alphadot \tilde\partial y_\mu - 4
 \tilde\thetabar
 \tilde\thetabar \tilde\partial \tilde\theta_\alpha\cr
 \bar q_\alphadot &= \bar p_\alphadot +i\theta^\alpha
 \sigma^\mu_{\alpha\alphadot} \partial x_\mu +{1\over 2}
 \theta\theta \partial\thetabar_ \alphadot -{3\over 2}
 \partial ( \thetabar_\alphadot\theta\theta) \cr
 \tilde {\bar q}_\alphadot &= \tilde{\bar p}_\alphadot +i\tilde
 \theta^\alpha \sigma^\mu_{\alpha\alphadot} \tilde\partial x_\mu
 +{1\over 2} \tilde \theta\tilde\theta\tilde \partial \tilde\thetabar
 _ \alphadot -{3\over 2}\tilde \partial ( \tilde\thetabar_\alphadot
 \tilde\theta\tilde\theta)   .}}

Since the vertex operator for a constant self-dual graviphoton
field strength is
 \eqn\gravip{\int d^2 z q_\alpha \tilde q_\beta F^{\alpha\beta}}
in the hybrid formalism, the action remains quadratic in this
background and there is no backreaction. The Lagrangian is
 \eqn\wslagIIaF{{\cal L}={1\over \alpha'}\left({1\over 2} \tilde
 \partial y^\mu \partial y_\mu -q_\alpha \tilde \partial
 \theta^\alpha
 + \bar d_\alphadot \tilde \partial\thetabar^\alphadot - \tilde
 q_\alpha \partial \tilde \theta^\alpha + \tilde{\bar d}_\alphadot
 \partial\tilde \thetabar^\alphadot + \alpha' F^{\alpha\beta} q_\alpha
\tilde q_\beta \right).
 }
The fields $q$ and $\tilde q$ can easily be integrated out using
their equations of motion
 \eqn\qtqeom{\eqalign{
 & \tilde \partial \theta^\alpha = \alpha' F^{\alpha\beta} \tilde
 q_\beta\cr
 &\partial\tilde \theta^\alpha = -\alpha' F^{\alpha\beta}
 q_\beta.}}

Let us examine the supersymmetries of the Lagrangian \wslagIIaF.
In addition to being invariant under the $Q_\alpha$ and $\tilde
Q_\alpha$ supersymmetries generated by
 \eqn\Qsup{\delta \theta^\alpha = \varepsilon^\alpha,\quad
 \delta \tilde\theta^\alpha = \tilde\varepsilon^\alpha,}
where $\varepsilon^\alpha$ and $\tilde \varepsilon^\alpha$ are
constants, the above Lagrangian is also invariant (up to total
derivatives) under the $\bar Q_\alphadot$ and $\tilde {\bar
Q}_\alphadot$ supersymmetries generated by
 \eqn\bQsup{\eqalign{
 &\delta \thetabar^\alphadot = - \bar\varepsilon^\alphadot \cr
 &\delta \tilde\thetabar^\alphadot = -
 \tilde{\bar\varepsilon}^\alphadot \cr
 &\delta y^\mu = 2i \sigma^\mu_{\alpha\alphadot}
 (\bar\varepsilon^\alphadot \theta^\alpha +
 \tilde{\bar\varepsilon}^\alphadot \tilde\theta^\alpha )\cr
 &\delta q_\alpha = 2i \bar\varepsilon^\alphadot
 \sigma_{\mu\alpha\alphadot} \partial y^\mu\cr
 &\delta \tilde q_\alpha = 2i \tilde{\bar\varepsilon}^\alphadot
 \sigma_{\mu\alpha\alphadot} \tilde\partial y^\mu\cr
  &\delta\theta^\alpha = 2i \alpha' F^{\alpha\beta}
 \tilde{\bar\varepsilon}^\alphadot y^\mu
 \sigma_{\mu\beta\alphadot}\cr
 &\delta\tilde\theta^\alpha = -2i\alpha'
 F^{\alpha\beta}{\bar\varepsilon}^\alphadot y^\mu
 \sigma_{\mu\beta\alphadot}.}}
These transformations are the stringy version of the supergravity
variation \news.  The last two transformations in \bQsup\
represent the deformation, and are needed because of the
deformation term $F^{\alpha\beta}q_\alpha\tilde q_\beta$ in the
Lagrangian. Unlike the other transformations in \bQsup, the last
two depend explicitly on $y^\mu$. This fact is similar to the
explicit $x$ dependence in
\news.  Therefore, the corresponding currents $\bar q_\alphadot$ and
$\tilde {\bar q}_\alphadot$ are not holomorphic and
antiholomorphic respectively.  They are similar to the currents of
the Lorentz symmetries in the worldsheet theory.

In order to establish that \bQsup\ are not only symmetries of the
worldsheet Lagrangian but are also symmetries of the string, we
should check that they commute with the BRST operators.  To do
that, we use $d_\alpha$ and $\tilde d_\alpha$ of \dqdef\ to show
that under \bQsup\
 \eqn\dtran{\delta d_\alpha = 8 (\tilde{\bar\varepsilon}_\alphadot
 \thetabar^\alphadot)\epsilon_{\alpha\gamma}\alpha' F^{\gamma\beta}
 d_\beta,\quad \delta\tilde d_\alpha = -8 ({\bar\varepsilon}_\alphadot
 \tilde\thetabar^\alphadot) \epsilon_{\alpha\gamma} \alpha'
 F^{\gamma\beta} \tilde d_\beta,}
where we have used the equations of motion \qtqeom.  Using \dtran\
it is clear that the BRST operators
 \eqn\brsto{\eqalign{
 &G = d_\alpha d^\alpha e^\rho,\quad \bar G = \bar d_\alphadot
 \bar d^\alphadot e^{-\rho}\cr
 &\tilde G =\tilde d_\alpha \tilde d^\alpha e^{\tilde\rho},\quad
 \tilde{\bar G} = \tilde{\bar d}_\alphadot \tilde{\bar d}^\alphadot
 e^{-\tilde\rho}, }}
are invariant.  This establishes that \bQsup\ are symmetries of
the theory.

One can easily read off the algebra of the deformed $\CN=2$ $d=4$
supersymmetry from the above transformations. Identifying
$(Q_\alpha, \tilde Q_\alpha)$ with $(Q_{\alpha 1},Q_{\alpha 2})$
and $(\bar Q_\alphadot,\tilde{\bar Q}_\alphadot)$ with $(\bar
Q_{\alphadot 2},-\tilde{\bar Q}_{\alphadot 1})$, one obtains the
algebra \newalg.

\newsec{$D$-branes}

Consider first the system without the background graviphoton. If
the worldsheet ends on a $D$-brane, the boundary conditions are
easily found by imposing that there is no surface term in the
equations of motion. For a boundary at $z=\tilde z$, we can use
the boundary conditions $\theta(z=\tilde z)=\tilde \theta(z=\tilde
z)$, $q(z=\tilde z)=\tilde q(z=\tilde z)$, etc. Then the solutions
of the equations of motion are such that $\theta(z)=\tilde
\theta(\tilde z)$, $q(z)=\tilde q(\tilde z)$, etc.; i.e.\ the
fields extend to holomorphic fields beyond the boundary.  The
boundary breaks half the supersymmetries preserving only $\oint q
dz + \oint \tilde q d\tilde z$ and $\oint \bar q dz + \oint \tilde
{\bar q} d\tilde z$.  From a spacetime point of view, these
unbroken supersymmetries are realized linearly on the brane, while
the other supersymmetries, which are broken by the boundary
conditions, are realized nonlinearly.

Now, let us consider the system with nonzero $F^{\alpha\beta}$.
After integrating out $q$ and $\tilde q$ the relevant part of the
worldsheet Lagrangian is
  \eqn\wslagIIaii{{\cal L}_{eff}= \left({1\over \alpha'^2 F}
  \right)_{\alpha\beta}  \partial\tilde \theta^\alpha \tilde
  \partial   \theta^\beta . }
The appropriate boundary conditions are
 \eqn\bouncondi{\theta^a (z=\tilde z) =\tilde \theta^a (z=
 \tilde z), \qquad \partial \tilde \theta ^a(z=\tilde z)
 = -\tilde \partial \theta^a(z=\tilde z).}
The first condition states that the superspace has half the number
of $\theta$s.  The second condition guarantees, using \qtqeom\
that $q_\alpha(z=\tilde z)=\tilde q_{\alpha}(z=\tilde z)$.

It is clear that the supercharges $\oint q dz + \oint \tilde q
d\tilde z$ are preserved by the boundary conditions and are still
realized linearly on the brane.  However, the supercharges $\oint
\bar q dz + \oint \tilde {\bar q} d\tilde z$ are broken and are
realized nonlinearly.

The propagator of $\theta$ is found to be
 \eqn\props{
 \langle \theta^\alpha(z,\tilde z) \theta^\beta(w,\tilde
 w)\rangle = {\alpha'^2F^{\alpha\beta} \over 2\pi i}
 \log{\tilde z-w\over z-\tilde w}}
with the branch cut of the logarithm outside the worldsheet.
Therefore for two points on the boundary $z=\tilde z=\tau$ and
$w=\tilde w= \tau'$
 \eqn\bounpro{\langle \theta^\alpha(\tau)\theta^\beta(\tau')
 \rangle = {\alpha'^2 F^{\alpha\beta} \over 2} {\rm sign}
 (\tau-\tau').}
Using standard arguments about open string coupling, this leads to
 \eqn\noncog{\{\theta^\alpha,\theta^\beta\}=\alpha'^2
 F^{\alpha\beta}=C^{\alpha\beta} \not=0;}
i.e.\ to a deformation of the anticommutator of the $\theta$s.  It
is important that since the coordinates $\bar \theta$ and $y$ were
not affected by the background coupling \gravip, they remain
commuting. In particular we derive that $[y^\mu,y^\nu]=
[y^\mu,\theta^\alpha]=0$, and therefore $[x^\mu,x^\nu]\not=0$ and
$[x^\mu,\theta^\alpha]\not=0$ \SeibergYZ.

For a partial list of other references on noncommuting $\theta$s
see \refs{\CasalbuoniHX\CasalbuoniBJ\CasalbuoniTZ
\BrinkNZ\SchwarzPF\FerraraMM\KlemmYU\AbbaspurXJ
\OoguriTT\KawaiYF\ChepelevGA\DavidKE-\BrittoAJ}.

\newsec{The zero slope limit}

The theory on the branes is simplified in the zero slope limit
$\alpha'\to 0$.  If we want to preserve the nontrivial
anticommutator of the $\theta$s \noncog, we should scale
 \eqn\zeros{\alpha'\to 0,\qquad F^{\alpha\beta}\to \infty,
 \qquad \alpha'^2 F^{\alpha\beta}=C^{\alpha\beta}={\rm fixed}.}
For $C=0$ the low energy field theory on the branes is
super-Yang-Mills theory.  For nonzero $C$ it is \SeibergYZ
 \eqn\caction{S= S_0 -i \int d^4 x C^{\mu\nu} \Tr( v_{\mu\nu}
 \bar\lambda\bar\lambda) +{1\over {4 }} \int d^4 x
 C^{\mu\nu}C_{\mu\nu} \Tr(\bar\lambda\bar\lambda)^2}
where $S_0 = \int d^4 x Tr[-{1\over 2}v^{\mu\nu}v_{\mu\nu}
-2i \lambda\sigma^m\bar\lambda +D^2]$
is the usual super-Yang-Mills action, $v^{\mu\nu}$ is
the Yang-Mills field strength, $\bar\lambda^\alphadot$ is the
antichiral gaugino, and $C^{\mu\nu} =
(\sigma^{\mu\nu})_{\alpha\beta} C^{\alpha\beta}$.

We will now examine the supersymmetries of the action \caction.

Under the $\CN={1\over 2}$ supersymmetries generated by $Q_\alpha
+ \tilde Q_\alpha$, the chiral superfield $W^\alpha$ transforms as
 \eqn\deW{\delta W^\beta = \eta_+^\alpha{\partial \over{\partial
 \theta^\alpha}} W^\beta}
where $\eta_+^\alpha$ is the constant supersymmetry parameter. In
component fields, these transformations are \SeibergYZ
 \eqn\stransf{\eqalign{
 &\delta\lambda= i\eta_+ D +\sigma^{\mu\nu}\eta_+
 (v_{\mu\nu} +{i\over 2}C_{\mu\nu}\bar\lambda\bar\lambda) \cr
 &\delta v_{\mu\nu}=i\eta_+ (\sigma_\nu \nabla_\mu -
 \sigma_\mu\nabla_\nu) \bar\lambda \cr
 &\delta D= -\eta_+\sigma^\mu \nabla_\mu\bar\lambda \cr
 &\delta\bar\lambda=0.}}
The action of \caction\ is invariant under these transformations
after including the $C_{\mu\nu}$-dependent terms in the
transformations and in the action.

However, under the $\CN={3\over 2}$ supersymmetries corresponding
to $(Q_\alpha -\tilde Q_\alpha, \bar Q_\alphadot, \tilde{\bar
Q}_\alphadot)$, the component fields transform nonlinearly since
these transformations do not preserve the $D$-brane boundary
conditions $\theta^\alpha= \tilde\theta^\alpha$ and
$\bar\theta^\alphadot=\tilde{\bar\theta}^\alphadot$.\foot{
Actually, since these supersymmetries are {\it spontaneously}
broken on the $D$-brane, the corresponding supercharges $(Q_\alpha
-\tilde Q_\alpha, \bar Q_\alphadot, \tilde{\bar Q}_\alphadot)$ do
not exist.  We will denote by $(Q_\alpha -\tilde Q_\alpha, \bar
Q_\alphadot, \tilde{\bar Q}_\alphadot)$ the transformations of the
supersymmetries.} To determine the nonlinear transformation of
these component fields, it is useful to recall that in a
background with Abelian chiral and antichiral superfields
$W^\alpha$ and $\bar W^\alphadot$, the $D$-brane boundary
conditions $\theta^\alpha -\tilde\theta^\alpha=0$ and
$\bar\theta^\alphadot - \tilde{\bar\theta}^\alphadot=0$ are
modified to $\theta^\alpha -\tilde\theta^\alpha={1\over 4} \alpha'
W^\alpha$ and $\bar\theta^\alphadot -\tilde{\bar\theta}^\alphadot
={1\over 4}\alpha'\bar W^\alphadot$ to leading order in $\alpha'$
\refs{\PershinBF,\BaggerWP,\BaggerPI}. Just as the bosonic
$D$-brane boundary conditions $\partial_\sigma x_\mu =0$ is
modified to $\partial_\sigma x_\mu = \alpha' (\p_\mu A_\nu-\p_\nu
A_\mu)\p_\tau x^\nu$ in the presence of the Maxwell vertex
operator $\int d\tau A_\mu (x) \partial_\tau x^\mu$
\refs{\FradkinFT,\YostYY}, the superstring $D$-brane boundary
conditions are modified to $\theta^\alpha
-\tilde\theta^\alpha={1\over 4} \alpha' W^\alpha$ and
$\thetabar^\alphadot -\tilde\thetabar^\alphadot={1\over 4} \alpha'
\bar W^\alphadot$ in the presence of the super-Maxwell vertex
operator which includes the term $\int d\tau (W^\alpha d_\alpha +
\bar W^{\alphadot} \bar d_\alphadot)$ \PershinBF. So by comparing
with the supersymmetry transformations of
$(\theta,\thetabar,\tilde\theta, \tilde{\bar\theta})$ in
\Qsup\bQsup, one learns the nonlinear transformations of the
Abelian gauginos $\lambda^\alpha$ and $\bar\lambda^\alphadot$ to
leading order in $\alpha'$. Note that the nonlinear
transformations of all other super-Yang-Mills fields are higher
order in $\alpha'$.

For example, under $Q_\alpha -\tilde Q_\alpha$, $\delta
(\theta-\tilde\theta) =\varepsilon-\tilde\varepsilon$ in \bQsup\
implies that the chiral Abelian gaugino transforms inhomogeneously
as
 \eqn\etao{\delta\lambda^\alpha ={{4i}\over {\alpha'}} (\varepsilon^\alpha -
 \tilde\varepsilon^\alpha)\equiv \eta_-^\alpha.}
This transformation clearly leaves invariant the action of
\caction.  In the zero slope limit we should take $\varepsilon -
\tilde\varepsilon \to 0$ with fixed $\eta_-$ in order to have a
finite answer.

Under $\bar Q_\alphadot - \tilde {\bar Q}_\alphadot$, the
antichiral Abelian gaugino transforms inhomogeneously as
 \eqn\etabo{\delta\bar\lambda^\alphadot = {{4i}\over { \alpha'}}
 (\bar\varepsilon^\alphadot -\tilde{\bar\varepsilon}^\alphadot)\equiv
 \bar \eta_-^\alphadot,}
and therefore in the zero slope limit we should take
$\bar\varepsilon -\tilde{\bar\varepsilon} \to 0$ with fixed $\bar
\eta_-$ in order to have a finite answer. In addition to \etabo,
since $\delta(\theta^\alpha+\tilde\theta^\alpha)=-{1\over 2} \bar
\eta_-^\alphadot C^{\alpha\beta} y_\mu
\sigma^\mu_{\beta\alphadot}$ is finite in the zero slope limit,
the chiral superfield $W^\gamma$ transforms as
 \eqn\vaWet{\delta W^\gamma = -{1\over 2}\bar \eta_-^\alphadot
 C^{\alpha\beta} y_\mu\sigma^\mu_{\beta\alphadot} {\partial
 \over{\partial\theta^\alpha}} W^\gamma.}
Therefore, to leading order in $\alpha'$, the component fields
transform as
 \eqn\snon{\eqalign{
 &\delta\lambda= i\eta(y,\bar \eta_-) D
 +\sigma^{\mu\nu}\eta(y,\bar \eta_-) (v_{\mu\nu}
 +{i\over 2} C_{\mu\nu}\bar\lambda\bar\lambda) \cr
 &\delta v_{\mu\nu}=i\eta(y,\bar \eta_-)
 (\sigma_\nu \nabla_\mu - \sigma_\mu\nabla_\nu) \bar\lambda \cr
 &\delta D= -\eta(y,\bar \eta_-) \sigma^\mu
 \nabla_\mu\bar\lambda\cr
 &\delta\bar\lambda= \bar \eta_-.}}
where $ \eta(y,\bar \eta_-)^\alpha \equiv -{1\over 2}\bar
 \eta_-^\alphadot C^{\alpha\beta} y_\mu
 \sigma^\mu_{\beta\alphadot}.$

To explicitly check that \caction\ is invariant under the
transformation of \snon, first note that under the inhomogeneous
transformation of $\bar\lambda$,
 \eqn\changeone{\delta_{inhomogeneous} S = -2i \int d^4 x
 C^{\mu\nu} \Tr[ v_{\mu\nu} \bar\eta_- \bar\lambda] + \int d^4 x
 C^{\mu\nu} C_{\mu\nu} \Tr[ (\bar\eta_-\bar\lambda)
 (\bar\lambda\bar\lambda)].}
And under the $y$-dependent transformation parameterized by
$\eta(y,\bar \eta_-)$ in \snon,
 \eqn\changetwo{\delta_{homogeneous} S = \int d^4 x J_\mu^\alpha
 {\partial\over{\partial y_\mu}}\eta(y,\bar \eta_-)_\alpha}
where $J_\mu^\alpha$ is the conserved supersymmetry current
associated with the invariance of \caction\ under the global
supersymmetry transformation of \stransf. One can easily compute
that
 \eqn\conscur{J_{\mu~\alpha} = \Tr[
 -2i v^{\rho\tau}(\sigma_{\rho\tau})_\alpha{}^\beta
\sigma_{\mu~\beta\gammadot}\bar\lambda^\gammadot
 + 2 C_{\mu\nu} \sigma^\nu_{\alpha\alphadot} \bar\lambda^\alphadot
 (\bar\lambda\bar\lambda )]}
and
 \eqn\derep{{\partial\over{\partial y^\mu}}\eta(y,\bar
 \eta_-)_\alpha= -{1\over 2}(\bar \eta_-)_\alphadot
 C_{\alpha\beta}\sigma_\mu^{\beta\alphadot}.}
Using \conscur\ and \derep\ in \changetwo, one finds that
\changetwo\ cancels \changeone\ so the action is invariant.

Lastly, under $\bar Q_\alphadot + \tilde {\bar Q}_\alphadot$, the
chiral Abelian gaugino transforms inhomogeneously as
 \eqn\lastt{\delta\lambda^\alpha = -8(\bar\varepsilon^\alphadot +
 \tilde{\bar\varepsilon}^\alphadot) (\alpha')^{-2} C^{\alpha\beta}
 y_\mu \sigma^\mu_{\beta\alphadot}\equiv -8\bar \eta_+^\alphadot
 C^{\alpha\beta} y_\mu \sigma^\mu_{\beta\alphadot}.}
Note that in this case we need $\bar\varepsilon^\alphadot +
\tilde{\bar\varepsilon}^\alphadot \sim \alpha'^2$ in order to have
a finite transformation in the zero slope limit. Since
$\delta(\thetabar^\alphadot+\tilde\thetabar^\alphadot)=
(\bar\varepsilon^\alphadot + \tilde{\bar\varepsilon}^\alphadot)
\sim \alpha'^2$, we should set it to zero in the zero slope limit.
Therefore, under this transformation \lastt\ is the only
variation. To check invariance of \caction\ under \lastt, note
that the inhomogeneous transformation of $\lambda^\alpha$ leaves
the action invariant since $\sigma^\mu_{\alpha\alphadot}
{\partial\over{\partial y^\mu}} \delta\lambda^\alpha=0$.

We conclude that the action \caction\ is invariant under the
linear $\CN={1\over 2}$ supersymmetry transformations
\deW\stransf\ with parameter $\eta_+$ and under the nonlinear
$\CN={3\over 2}$ supersymmetry transformations \etao\snon\lastt\
with parameters $\eta_-$, $\bar \eta_-$, $\bar \eta_+$.

It would be interesting to extend our analysis to higher-order
terms in $\alpha'$.  This amounts to finding a noncommutative
superspace generalization to the $d=4$ supersymmetric Born-Infeld
action. Just as the standard supersymmetric Born-Infeld action
realizes $\CN=1+1$ supersymmetry (four supercharges are realized
linearly and four supercharges are realized nonlinearly)
\refs{\BaggerWP,\BaggerPI}, the noncommutative superspace
Born-Infeld theory should realize $\CN={1\over 2} + {3\over 2}$
supersymmetry.

\bigskip
\centerline{\bf Acknowledgements}

It is a pleasure to thank H. Ooguri, C. Vafa and E. Witten for
helpful discussions. The work of NS was supported in part by DOE
grant \#DE-FG02-90ER40542 and NSF grant PHY-0070928 to IAS. NB
would like to thank the IAS for its hospitality where part of this
work was done, and FAPESP grant 99/12763-0, CNPq grant 300256/94-9
and Pronex grant 66.2002/1998-9 for partial financial support.

\listrefs

\end